\documentclass[a4paper]{article}
\usepackage{aqis}
\usepackage{hyperref}
\usepackage{graphicx}
\usepackage{qcircuit}
\usepackage{float}
\usepackage{stfloats}
\usepackage[utf8]{inputenc}
\usepackage{graphicx,overpic}
\begin{document}

\title{
   Classification of Financial Data Using Quantum Support Vector Machine 
}

\author{
 Seemanta Bhattacharjee
   \affiliation{1}
  \email{babune99@gmail.com}
  \and
 MD. Muhtasim Fuad
   \affiliation{1}
  \email{muhtasimfuad625@gmail.com}
  \and
  A.K.M. Fakhrul Hossain
  \affiliation{1}
  \email{a.k.m.fakhrul.hossain@gmail.com}
  }
  
\address{1}{
  Shahjalal University Of Science and Technology, Sylhet-3114, Bangladesh
}

\abstract{
Quantum Support Vector Machine is a kernel-based approach to classification problems. We study the applicability of quantum kernels to financial data, specifically our self-curated Dhaka Stock Exchange (DSEx) Broad Index dataset. To the best of our knowledge, this is the first systematic study of quantum kernels applied to this dataset. Working within the empirical quantum advantage (EQA) framework of Krunic et al., we benchmark several quantum kernels against a classical RBF-kernel SVM baseline, propose the best-performing kernel for this dataset, and relate the observations to the Phase Space Terrain Ruggedness Index metric. We estimate the resources needed to carry out these investigations on a larger scale for future practitioners.  
}


\maketitle


\section{Introduction}
Quantum Computers are expected to benefit the finance industry significantly with the ability to solve specific problems more quickly and with greater accuracy than the best-known classical approaches. While classical machine learning techniques, such as support vector machines (SVMs), have proven highly effective in tracking the stock market and optimizing stock option purchases while minimizing risk, they have limitations in learning some of the most intricate patterns \cite{shen2012stock}. This motivates the exploration of quantum feature maps, which have shown promising results in outperforming classical SVMs on certain problem domains. For instance, a recent study demonstrated the advantage of a quantum feature map over classical SVMs on an artificial dataset based on the discrete logarithm problem (DLP) \cite{kristantemme2020}.

Classical SVMs may struggle to handle the high-dimensional, non-linear, and noisy data often found in stock market datasets. Additionally, stock market data exhibit non-stationary behavior, with the statistical properties of the data changing over time, making this problem a prime candidate for quantum algorithms. For this experiment, we have curated a Dhaka Stock Exchange Broad Index dataset to predict daily changes in the index. Quantum feature maps \cite{schuld2019quantum, havlivcek2019supervised} are used to build quantum kernels and to compare their performance in this data set with their classical counterparts and among themselves. To predict the configuration spaces where quantum kernels may provide an advantage over classical approaches, the study uses the Phase Space Terrain Ruggedness Index (PTRI) \cite{shehab2022} as a global metric. Our experiment aligns with the findings of this metric. We also provide insights on estimating the quantum resources required to conduct this experiment.

This study serves as a framework that will be useful to future practitioners, resembling clinical trials. We primarily use small-sized datasets. This choice is intentional, as small datasets often present a more challenging classification problem closer to the real-world scenarios faced by financial institutions and market analysts. We present the first systematic study of QSVM applied to the DSEx Broad Index dataset and provide a comparison of various quantum kernels for financial data classification.

\section{Results and Discussion}
This study focuses on a binary classification problem using the Dhaka Stock Exchange Broad Index dataset. We conduct experiments on both classical and quantum methods simultaneously across various data sizes. This work provides a robust comparison of several quantum kernels for this dataset type. 
Balanced Accuracy and F1 Score are used to measure the kernels' performance for this research. Both of these metrics are commonly used in machine learning experiments, and since we randomly selected subsets of data to train and test the kernels, these metrics provide a reliable way to measure performance.
The results of this study show that in most configurations, quantum kernels outperform the classical RBF-kernel SVM baseline (see Methods) in predicting daily changes in the Dhaka Stock Exchange Broad Index.

\begin{figure}[H]
    \centering
    \includegraphics[width=0.95\linewidth]{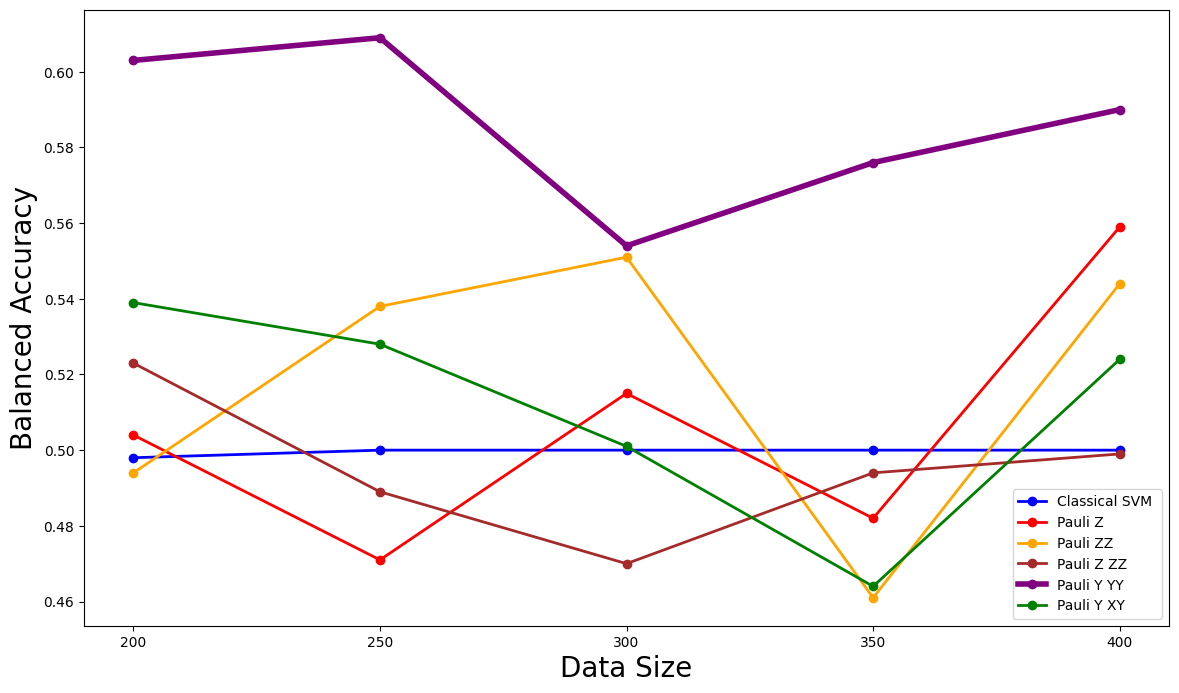}
    \caption{Comparing average Balanced Accuracy of quantum kernels with classical SVM on datasets where classical SVM performance is closest to mean performance, across varying dataset sizes and 7 features.}
    \label{fig:multiplefeaturemapsba}
\end{figure}

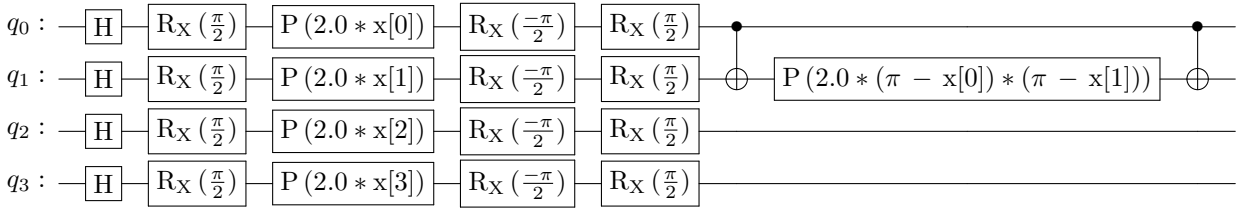
\begin{figure*}[t]
\centering
\[
\Qcircuit @C=1.0em @R=0.2em @!R { \\
	 	\nghost{{q}_{0} :  } & \lstick{{q}_{0} :  } & \gate{\mathrm{H}} & \gate{\mathrm{R_X}\,(\mathrm{\frac{\pi}{2}})} & \gate{\mathrm{P}\,(\mathrm{2.0*x[0]})} & \gate{\mathrm{R_X}\,(\mathrm{\frac{-\pi}{2}})} & \gate{\mathrm{R_X}\,(\mathrm{\frac{\pi}{2}})} & \ctrl{1} & \qw & \ctrl{1} & \qw\\
	 	\nghost{{q}_{1} :  } & \lstick{{q}_{1} :  } & \gate{\mathrm{H}} & \gate{\mathrm{R_X}\,(\mathrm{\frac{\pi}{2}})} & \gate{\mathrm{P}\,(\mathrm{2.0*x[1]})} & \gate{\mathrm{R_X}\,(\mathrm{\frac{-\pi}{2}})} & \gate{\mathrm{R_X}\,(\mathrm{\frac{\pi}{2}})} & \targ & \gate{\mathrm{P}\,(\mathrm{2.0*(\pi\,-\,x[0])*(\pi\,-\,x[1])})} & \targ & \qw\\
	 	\nghost{{q}_{2} :  } & \lstick{{q}_{2} :  } & \gate{\mathrm{H}} & \gate{\mathrm{R_X}\,(\mathrm{\frac{\pi}{2}})} & \gate{\mathrm{P}\,(\mathrm{2.0*x[2]})} & \gate{\mathrm{R_X}\,(\mathrm{\frac{-\pi}{2}})} & \gate{\mathrm{R_X}\,(\mathrm{\frac{\pi}{2}})} & \qw & \qw & \qw & \qw\\
	 	\nghost{{q}_{3} :  } & \lstick{{q}_{3} :  } & \gate{\mathrm{H}} & \gate{\mathrm{R_X}\,(\mathrm{\frac{\pi}{2}})} & \gate{\mathrm{P}\,(\mathrm{2.0*x[3]})} & \gate{\mathrm{R_X}\,(\mathrm{\frac{-\pi}{2}})} & \gate{\mathrm{R_X}\,(\mathrm{\frac{\pi}{2}})} & \qw & \qw & \qw & \qw\\
\\ }
\]
\caption{Cropped circuit of Pauli Y YY Feature Map for Quantum Kernel Calculation.}
    \label{fig:quantum_circuit}
\end{figure*}


From figure \ref{fig:multiplefeaturemapsba}, we observe that the quantum kernel built using the Pauli Y YY feature map is the most suitable for this dataset. We notice that the kernels built using the Pauli Y YY feature map consistently outperformed all other quantum kernels and the classical Support Vector Machine (SVM) baseline for every point in the feature-dataset configuration space. The experiments run on 200 to 400 data points and use 5 to 7 features. 

For further investigation, we run the same experiments on datasets where the performance of classical SVMs was closest to their maximum and minimum performance, respectively. Our findings show that the kernel built using the Pauli Y YY feature map outperforms classical SVMs.

The Pauli Y YY feature map performs consistently better in every configuration we tested, whereas the classical SVM performs worse than its average. While comparing the position of the data points with the horizontal 'zero-advantage line' in Figure 3, we observe empirical quantum advantage (EQA), in the sense of \cite{shehab2022}, for all problem instances in the configuration space.


\begin{figure}[H]    \centering
    \begin{overpic}[width=\linewidth]{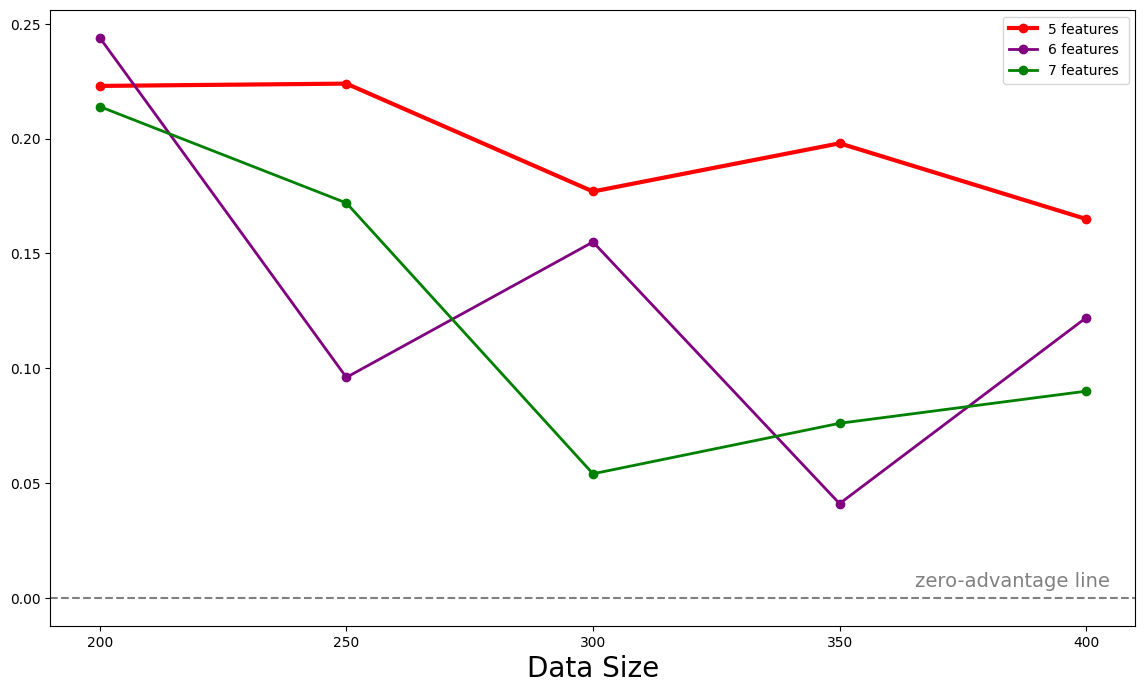}
        \put(-4, 20){\rotatebox{90}{\scriptsize Balanced Accuracy}}
    \end{overpic}
    \caption{Difference of average Balanced Accuracy of classical SVM vs Pauli Y YY Kernel on datasets where classical SVM performance is closest to minimum performance, across varying sizes and 5 to 7 number of features.}
    \label{fig:multiplekernelsMINba}
\end{figure}

\begin{figure*}[h]
    \centering
    \includegraphics[width=\linewidth]{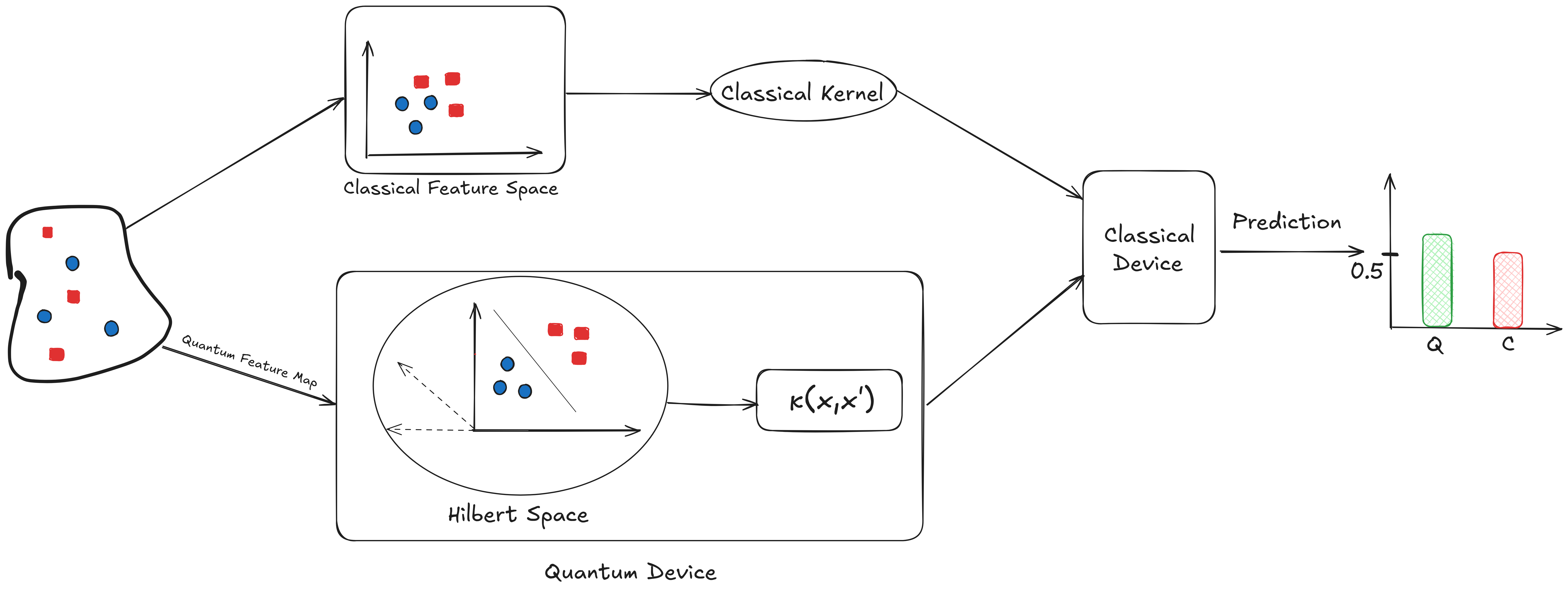}
    \caption{Illustration of the Classical and Quantum classification approaches.}
    \label{fig:approaches}
\end{figure*}

Figure \ref{fig:multiplekernelsMINba} implies the impact and importance of the quantum approach in real-world scenarios where we face the scarcity of stock market data and where the classical SVM performs worse than the average performance.

Regarding the resources required for quantum kernel classification, we can potentially minimize the number of qubits required to perform experiments with fixed features using Bloch Sphere Encoding \cite{heredge2021}. This approach has the potential to reduce the resources needed for these experiments. Additionally,  exploring the impact of different hyperparameter settings and regularization techniques on the algorithm's performance can further improve its accuracy and generalizability. 

Our comparisons should be read alongside the scale of the study: a single random seed, an RBF-kernel classical baseline without extensive hyperparameter tuning, and predominantly noiseless simulation. While this study used small datasets to thoroughly explore the intrinsic capabilities of quantum kernels, future investigations would benefit from using larger and more diverse financial datasets to further validate the generalizability of our findings. Expanding the scope of the data sources could provide additional insights into the practical applicability of quantum machine learning techniques in the finance domain.


\section{Methods}

\subsection*{Quantum Feature Map}

An important step in achieving a quantum advantage over classical methods involves creating feature maps derived from quantum circuits with significant simulation challenges for classical computers. Pauli Y YY generates a quantum feature map of the form:
\begin{equation*}
\resizebox{\columnwidth}{!}{$
\mathcal{U}_\Phi(\mathbf{x}) = \left( \exp\left(i \sum_{jk} \phi_{\{j,k\}}(\mathbf{x}) Y_j \otimes Y_k\right) \exp\left(i \sum_j \phi_{\{j\}}(\mathbf{x}) Y_j\right) H^{\otimes n} \right)^d 
$}
\end{equation*}

The unitary operation $\mathcal{U}_\Phi(\mathbf{x})$ encodes classical input $\mathbf{x}$ into a quantum state using a parameterized quantum circuit. It consists of alternating layers of Hadamard gates ($H^{\otimes n}$), single-qubit rotations parameterized by $\phi_{\{j\}}(\mathbf{x})$ with Pauli-$Y$ operators, and entangling two-qubit interactions parameterized by $\phi_{\{j,k\}}(\mathbf{x})$. This is from a family of feature maps conjectured to be hard to simulate classically \cite{havlivcek2019supervised} and that can be implemented as short-depth circuits on near-term quantum devices. Figure \ref{fig:quantum_circuit} illustrates a sample circuit computing the feature map of a dataset with four features. We also compare the results separately using Pauli Z, Pauli ZZ, Pauli Y ZZ, and Pauli Z ZZ feature maps for every classification problem.

\subsection*{DSEx Dataset}

The DSEx dataset is curated through a comprehensive collection of features from various online sources, which are systematically merged to form a unified dataset. Exploratory Data Analysis is conducted on the dataset to provide initial insights. More economic features are added that usually directly impact the change of the stock market's trend.

The DSEx index analyzes the performance of the top $20\%$ of listed firms in terms of market capitalization and liquidity, and it is computed using the component stocks' weighted average market capitalization. It is used as a benchmark to assess the overall performance. The historical data of the Dhaka Stock Exchange Broad was sourced from Investing.com \cite{investing2024}. In addition, we obtained historical data of gold prices spanning a decade from usagold.com \cite{usagold2024}. The curated dataset was created by merging the two datasets, resulting in a total of 460 data points.

\subsection*{Quantum Experiment}
Using Qiskit and IBM Quantum simulators and actual hardware, classical SVM models are trained based on radial basis function kernels and quantum models using custom kernels.
The kernel is then saved and used to train the SVM model using a precomputed option in scikit-learn on the classical computer. To directly predict the labels, we utilize the predict function from scikit-learn. This procedure is susceptible to variation based on the random seed given to the svm.SVC function. To ensure consistency, calculations and comparisons only use one random value. The quantum simulations are performed without the use of noise models on $qasm\_simulator$. Figure \ref{fig:approaches} illustrates how the data flows during both the quantum and classical approaches.

These experiments are almost an order of magnitude less than previous large-scale quantum machine learning studies, with no circuit executed more than 1024 times (i.e., a maximum of 1024 shots) \cite{peters2021machine}. However, this contributes to increased sampling noise.

\subsection*{IBM Quantum Hardware}
$ibm\_nairobi$ was a 7-qubit Quantum Processing Unit (QPU) which was available on IBM Quantum Services. Multiple instances of our experiment ran on IBM quantum computing hardware, and the results consistently align with those obtained from simulations.

\section{PTRI Verification}

\begin{figure}[H]
    \centering
    \includegraphics[width=0.9\linewidth]{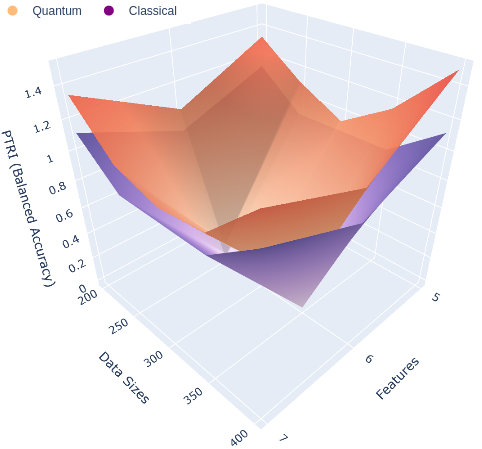}
    \caption{PTRI scores for classical SVM and QSVM in a $15$ point configuration space.}
    \label{fig:ptri}
\end{figure}

PTRI systematically identifies machine learning problems where quantum kernels may exhibit empirical superiority \cite{shehab2022}. A geophysics-inspired strategy is proposed to delineate regions of potential Empirical Quantum Advantage (EQA) within datasets to facilitate the selection of the subset of problems that could benefit from a quantum kernel. To this end, one viable strategy involves analyzing the ruggedness of the manifold via PTRI. To ascertain the PTRI values for the entire configuration space, we evaluated the balanced accuracy metric and represented the results graphically in Figure \ref{fig:ptri}. 

In figure \ref{fig:ptri}, we present a comparison of PTRI scores, with the purple-colored surface depicting Classical computing, and the peach-colored surface representing Quantum computing. 


The data points for each configuration space coordinate were averaged over two selected datasets, resulting in a total of $15$ points for each plotted configuration space. The z-axis depicts the metric, while the x- and y-axes represent the number of features and data size, respectively. We verified the PTRI metric for this dataset and observed that classical SVM’s balanced accuracy did not fluctuate significantly, even as the size of the training set increased. We noted that quantum kernels performed better on smoother terrain.



\section{Resource Estimation}


A thorough assessment of the resources required to construct the circuit essential for developing the quantum kernel using the Y YY feature map is conducted. 

Our findings provide a framework for future researchers to estimate resources for conducting similar experiments more comprehensively. The number of gates and circuit depth increases as the number of features or circuit repetition increases. 

The total number of gates needed in a quantum experiment can be calculated by the following equation, where $F$ defines the number of features and $R$ defines the circuit's repetition number.

 $$\text{Total} = (11 \times F - 7) \times R$$
   
The circuit consists of four different types of gates which are $ H, R_x, P \text{ and } C_x$ gates. The total number of $ H, R_x, P \text{ and } C_x$ gates needed individually can be found by the following equations respectively: 
    $$H = F \times R$$
    $$R_x = (6 \times F - 4) \times R$$
    $$P = (2 \times F - 1) \times R$$
    $$C_x = (2 \times F - 2) \times R$$

We also provide the equation to calculate the quantum circuit depth. The depth of a quantum circuit can be calculated using the following equation:
    
    $$\text{Depth} = (5 \times F - 1) \times R$$
 
The number of qubits required is independent of the number of repetitions in the circuit. The required number of qubits precisely matches the number of features, regardless of the number of repetitions.





\section{Variability and Errors}
\begin{figure}[H]
    \centering
    \includegraphics[width=\linewidth]{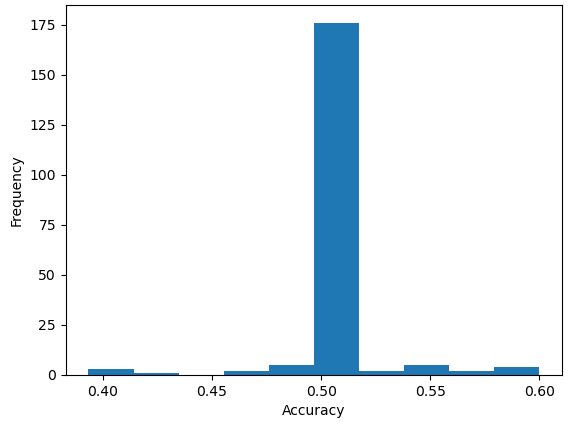}
    \caption{Examining Balanced Accuracy Variability on Configuration Space Coordinate with 200 Samples and 5 Features through 200 different experiments.}
    \label{fig:variability}
\end{figure}

The visualization of how the balanced accuracy is distributed for the Classical model is illustrated in the figure \ref{fig:variability}.


To illustrate the distribution of balanced accuracy in classical SVM, 200 separate experiments are run where the data size is 200 and number of features is 5. It shows a 2.1\% standard deviation while predicting the trend in classical domain which makes this a prime candidate for the application of quantum algorithms.

\section{Conclusion}

To our knowledge, this study is the first systematic application of quantum kernel methods to the Dhaka Stock Exchange Broad Index. The results show the potential of gaining an empirical quantum advantage over classical baselines on complex real-world datasets.

We utilize an end-to-end framework to document empirical quantum advantage on financial datasets. In the future, we can study the effects of classically hard-to-simulate kernels on these types of data. Practitioners can uncover future research directions that may improve the performance of quantum kernels for these types of datasets. Moreover, future research may also open doors to developing custom feature maps specifically designed for financial data characteristics. 

In this work, we present a small-scale case study of applying quantum machine learning to financial data. As quantum hardware continues to improve, larger and more rigorous studies will clarify how far these techniques can benefit the finance industry.

\vspace{1cm} 

\begin{center}
\subsection*{Acknowledgements}
\end{center}

The authors would like to thank Dr$.$ Omar Shehab and Sristy Sangskriti for their constant support and supervision throughout the process.


\end{document}